\documentclass[a4paper,12pt]{article}

\newcommand{\sect}[1]{\setcounter{equation}{0}\section{#1}}

\begin{document}
\topmargin 0pt \oddsidemargin 0mm

\renewcommand{\thefootnote}{\fnsymbol{footnote}}
\begin{titlepage}

\vspace{2mm}
\begin{center}
{\Large \bf First Law of Thermodynamics and Friedmann Equations of
Friedmann-Robertson-Walker Universe}
 \vspace{12mm}

{\large Rong-Gen Cai\footnote{e-mail address: cairg@itp.ac.cn}}

\vspace{5mm} {\em  Institute of Theoretical Physics, Chinese
Academy of Sciences, \\
 P.O. Box 2735, Beijing 100080, China \\
 CASPER, Department of Physics, Baylor University, Waco, TX76798-7316, USA}

 \vspace{5mm}
 {\large Sang Pyo Kim\footnote{e-mail address: sangkim@kunsan.ac.kr}}\\
 \vspace{5mm}
 {\em Department of Physics, Kunsan National University, Kunsan 573-701,
 Korea}\\

\end{center}

\vspace{45mm} \centerline{{\bf{Abstract}}} \vspace{5mm}
Applying
the first law of thermodynamics to the apparent horizon of a
Friedmann-Robertson-Walker universe and assuming the geometric
entropy given by a quarter of the apparent horizon area, we derive
the Friedmann equations describing the dynamics of the universe
with any spatial curvature. Using entropy formulae for the static
spherically symmetric black hole horizons in Gauss-Bonnet gravity
and in more general Lovelock gravity, where the entropy is not
proportional to the horizon area, we are also able to obtain the
Friedmann equations in each gravity theory. We also discuss some
physical implications of our results.
\end{titlepage}

\newpage
\renewcommand{\thefootnote}{\arabic{footnote}}
\setcounter{footnote}{0} \setcounter{page}{2}

\sect{Introduction}

Quantum mechanics together with general relativity predicts that a
black hole behaves like a black body, emitting thermal radiations,
with a temperature proportional to its surface gravity at the
black hole horizon and with an entropy proportional to its horizon
area \cite{Haw,Bek}. The Hawking temperature and horizon entropy
together with the black hole mass obey the first law of
thermodynamics \cite{firstlaw}. The formulae of black hole entropy
and temperature have a certain universality in the sense that the
horizon area and surface gravity are purely geometric quantities
determined by the spacetime geometry, once Einstein equations
determine the spacetime geometry.

Since the discovery of black hole thermodynamics in 1970's,
physicists have been speculating that there should be some
relation between black hole thermodynamics and Einstein equations.
Otherwise, how does general relativity know that the horizon area
of black hole is related to its entropy and the surface gravity to
its temperature \cite{Jac}? Indeed, Jacobson \cite{Jac} was able
to derive Einstein equations from the proportionality of entropy
to the horizon area together with the fundamental relation $\delta
Q= TdS$, assuming the relation holds for all local Rindler causal
horizons through each spacetime point. Here $\delta Q$ and $T$ are
the energy flux and Unruh temperature seen by an accelerated
observer just inside the horizon. On the other hand, Verlinde
\cite{Ver} found that for a radiation dominated
Friedmann-Robertson-Walker (FRW) universe, the Friedmann equation
can be rewritten in the same form as the Cardy-Verlinde formula,
the latter being an entropy formula for a conformal field theory.
Note that the radiation can be described by a conformal field
theory. Therefore, the entropy formula describing the
thermodynamics of radiation in the universe has the same form as
that of the Friedmann equation, which describes the dynamics of
spacetime. In particular, when the so-called Hubble entropy bound
is saturated, these two equations coincide with each other (for
more or less a complete list of references on this topic see, for
example, \cite{cm}). Therefore, Verlinde's observation further
indicates some relation between thermodynamics and Einstein
equations.

In a four dimensional de Sitter space with radius $l$, there is a
cosmological event horizon. This horizon, like a black hole
horizon, is associated with thermodynamic properties~\cite{GH}:
the Hawking temperature $T$ and entropy $S$,
\begin{equation}
 \label{1eq1}
 T = \frac{1}{2\pi l}, \ \ \ S = \frac{A}{4G},
\end{equation}
where $A= 4\pi l^2$ is the cosmological horizon area and $G$ is
the Newton constant. For an asymptotic de Sitter space such as a
Schwarzschild-de Sitter space, there still exists the cosmological
horizon, for which the area law of the entropy holds $S=A/4G$,
where $A$ denotes the cosmological horizon area, and whose Hawking
temperature is given by $T=\kappa/2\pi$, where $\kappa$ is the
surface gravity of the cosmological horizon. Suppose that some
matter with energy $dE$ passes through the cosmological horizon,
one then has
 \begin{equation}
 \label{1eq2}
 -dE=TdS.
 \end{equation}
It is easy to verify that the cosmological horizon in the
Schwarzschild-de Sitter space satisfies the relation (\ref{1eq2})
(see, for example, \cite{Cai1}). Equation (\ref{1eq2}) is just the
first law of thermodynamics for the cosmological horizon.

In the slow-roll inflationary model, the spacetime is a quasi-de
Sitter one. If the inflation period is followed by a flat universe
with radiation and dust matter, as in the standard big bang
universe, the cosmological event horizon is absent in such a
universe. However, an apparent horizon always exists, where the
expansion vanishes. Frolov and Kofman \cite{FK} employed the
approach proposed by Jacobson~\cite{Jac} to a quasi-de Sitter
geometry of inflationary universe, where they calculated the
energy flux of a background slow-roll scalar field (inflaton)
through the quasi-de Sitter apparent horizon and used the relation
(\ref{1eq2}). Although the topology of the local Rindler horizon
in Ref. \cite{Jac} is quite different from that of the quasi-de
Sitter apparent horizon considered in Ref. \cite{FK},  it was
found that this thermodynamic relation reproduces one of the
Friedmann equations with the slow-roll scalar field. It is assumed
in their derivation that
\begin{equation}
\label{1eq3}
 T=\frac{H}{2\pi}, \ \ \ S= \frac{\pi}{GH^2},
\end{equation}
where $H$ is a slowly varying Hubble parameter.

More recently, following Refs. \cite{Jac,FK}, Danielsson
\cite{Dan} has been able to obtain the Friedmann equations, by
applying the relation $\delta Q=TdS$ to a cosmological horizon to
calculate the heat flow through the horizon in an expanding
universe in an acceleration phase and assuming the same form for
the temperature and entropy of the cosmological horizon as those
in Eq. (\ref{1eq3}). Furthermore, Bousso \cite{Bousso} has
recently considered thermodynamics in the Q-space (quintessence
dominated spacetime). Because the equation of state of
quintessence is in the range of $-1 < \omega <-1/3$, the universe
undergoes an accelerated expansion, and thus the cosmological
event horizon exists in  the Q-space. However, Bousso argued that
a thermodynamic   description of the horizon is approximately
valid and thus it   would not matter much whether one uses the
apparent horizon or the event horizon. Indeed, for the Q-space,
the apparent horizon  radius $R_A$ differs from the event horizon
radius $R_E$ only by a small quantity: $R_A/R_E=1-\epsilon$, where
$\epsilon = 3(\omega +1)/2$. Using the relations
  \begin{equation}
  \label{1eq4}
  T = \frac{1}{2\pi R_A}, \ \ \  S = \frac{\pi R^2_A}{G},
  \end{equation}
Bousso showed that  the first law (\ref{1eq1}) of thermodynamics
holds at the apparent horizon of the Q-space.  While these authors
\cite{FK,Dan,Bousso} dealt with different  aspects of the relation
between the first law of thermodynamics and Friedmann equations,
they considered only a flat FRW universe.

There is a subtlety that the apparent horizon, the Hubble horizon
and the cosmological event horizon cannot be distinguished clearly
in some cases where one uses Eqs. (\ref{1eq3}), (\ref{1eq4}) or
more general forms $T=\kappa/2\pi$ and $S=A/4G$. Therefore, one
cannot be sure whether the first law holds for one of the
cosmological horizon, the Hubble horizon and the apparent horizon
or for some of them or for all of them. Further, when the spatial
curvature does not vanish, it would be an interesting issue to see
whether one can still derive or not the Friedmann equations from
the first law of thermodynamics and to check the relation between
thermodynamics and Einstein equations in a more general context.
In addition, it is well known that the area formula of black hole
entropy holds only in Einstein theory, that is, when the action of
gravity theory includes only a linear term of scalar curvature.
Therefore, it is worthwhile to study whether, given a relation of
entropy and horizon area, one can obtain the Friedmann equations
in the corresponding gravity theory from the first law of
thermodynamics. In this paper we are going to discuss these
issues.

This paper is organized as follows. In Sec.~2, we shall review and
clarify the discussions on the relation between thermodynamics and
Einstein equations in Refs. \cite{FK,Dan,Bousso} and extend the
relation to the case of a FRW universe with any spatial curvature
in arbitrary dimensions. Applying the first law of thermodynamics
to the apparent horizon and assuming the proportionality of
entropy and horizon area, we shall successfully derive the
Friedmann equations for the universe. In Sec.~3, we shall obtain
the Friedmann equations in Gauss-Bonnet gravity by employing the
entropy formula for static spherically symmetric Gauss-Bonnet
black holes. In Sec. 4, we shall discuss a more general case
within the Lovelock gravity. Finally, in Sec.~5, we shall discuss
physical implications of the relation.

\sect{Friedmann equations in Einstein gravity}

Let us start with an $(n+1)$-dimensional FRW universe with the
metric
\begin{equation}
\label{2eq1}
 ds^2= -dt^2 + a^2(t) \left( \frac{dr^2}{1-kr^2} + r^2
 d\Omega_{n-1}^2\right),
 \end{equation}
 where $d\Omega_{n-1}^2$ denotes the line element of an
 $(n-1)$-dimensional unit sphere and the spatial curvature constant
 $k=+1$, $0$ and $-1$ correspond to a closed, flat and open universe,
 respectively. The metric (\ref{2eq1}) can be rewritten
 as~\cite{BR}
 \begin{equation}
 \label{2eq2}
 ds^2 = h_{ab}dx^adx^b +\tilde r^2 d\Omega_{n-1}^2,
 \end{equation}
 where $\tilde r = a(t) r$ and $x^0=t$, $x^1=r$ and the
 2-dimensional metric $h_{ab} = {\rm diag} (-1, a^2/(1-kr^2))$.
 The dynamical apparent horizon, a marginally trapped surface with
 vanishing expansion, is determined by the
 relation $h^{ab} \partial_a \tilde r
 \partial _b \tilde r=0$. A simple calculation yields the radius
 of the apparent horizon
 \begin{equation}
 \label{2eq3}
 \tilde r_A = \frac{1}{\sqrt{H^2 +k/a^2}},
 \end{equation}
 where $H$ denotes the Hubble parameter, $H\equiv
 \dot a /a$. Here and hereafter the dots will represent derivatives with
 respect to the cosmic time $t$ in the metric (\ref{2eq1}). It can be
seen from (\ref{2eq3}) that when $k=0$, namely, for a flat
universe, the radius $\tilde r_A$ of the apparent horizon  has the
same value as the radius $\tilde r_H$ of the Hubble horizon, which
is defined as the inverse of the Hubble parameter, that is, $
\tilde r_H =1/H$. On the other hand, the cosmological event
horizon defined by
\begin{equation}
\label{2eq4}
 \tilde r_E = a(t) \int^{\infty}_{t}\frac{dt}{a(t)},
 \end{equation}
exists only for an accelerated expanding universe. As a
consequence, for a pure de Sitter universe with $k=0$, the
apparent horizon, the Hubble horizon and the cosmological event
horizon have the same constant value $1/H$. Note that though the
cosmological event horizon does not always exist for all FRW
universes, the apparent horizon and the Hubble horizon always do
exist. For a dynamical spacetime, the apparent horizon has been
argued to be a causal horizon and is associated with the
gravitational entropy and surface gravity \cite{Hay,BR}. Thus, for
our purpose, it would be convenient to employ the apparent horizon
and to apply the first law to the apparent horizon.

Following Refs. \cite{BR,Hay}, we define the work density by
\begin{equation}
\label{2eq5}
 W = -\frac{1}{2}T^{ab}h_{ab},
 \end{equation}
 and the energy-supply vector by
 \begin{equation}
 \label{2eq6}
 \Psi_a = T_a^{\ b} \partial _b \tilde r + W \partial_a \tilde r,
 \end{equation}
 where $T^{ab}$ is the projection of the $(n+1)$-dimensional
 energy-momentum tensor $T^{\mu\nu}$ of a perfect fluid matter
 in the FRW universe in the normal direction of the $(n-1)$-sphere.
 The work density at the apparent horizon should be regarded as
 the work done by a change of the apparent horizon, while the
 energy-supply at the horizon is the total energy flow through the
 apparent horizon. Then it is shown that one has \cite{Hay,BR}
 \begin{equation}
 \label{2eq7}
 \nabla E = A \Psi + W \nabla V,
 \end{equation}
 where $A = n \Omega_n \tilde r^{n-1}$ and  $V=
 \Omega_n \tilde r^n$ are the area and volume of an $n$-dimensional
  space with radius $\tilde r$, $\Omega_n=\pi ^{n/2}/\Gamma(n/2 +1)$
 being the volume of an $n$-dimensional unit ball, and that the total energy
  inside the space with radius $\tilde r$ is defined by
  \begin{equation}
  E = \frac{n(n-1)\Omega_n}{16\pi G}\tilde
  r^{n-2}(1-h^{ab}\partial_a\tilde r\partial_b \tilde r).
  \end{equation}
  The equation (\ref{2eq7}) is dubbed {\it the unified first
  law}~\cite{Hay}. According to thermodynamics, the entropy is
  associated with heat flow as $\delta Q=TdS$, and the heat flow
  is related to the change of energy of the given system.
  As a consequence, the
  entropy is finally associated with the energy-supply term. The
  latter can be rewritten as
  \begin{equation}
  \label{2eq9}
  A \Psi=\frac{\kappa}{8\pi G} \nabla A +\tilde
  r^{n-2}\nabla(\frac{E}{\tilde r^{n-2}}),
  \end{equation}
  where $\kappa$ is the surface gravity defined as
  \begin{equation}
  \label{2eq10}
  \kappa =\frac{1}{2\sqrt{-h}}\partial_a
  (\sqrt{-h}h^{ab}\partial_b\tilde r).
  \end{equation}
On the apparent horizon the last term in Eq. (\ref{2eq9})
vanishes, and one can then assign an entropy $S= A/4G$ to the
apparent horizon.

Next, we turn to calculating the heat flow $\delta Q$ through the
apparent horizon during an infinitesimal time interval $dt$. Heat
is one of forms of energy. Therefore, the heat flow $\delta Q$
through the apparent horizon is just the amount of energy crossing
it during that time internal $dt$. That is, $\delta Q=-dE$ is the
change of the energy inside the apparent horizon. Suppose that the
energy-momentum tensor $T_{\mu\nu}$ of the matter in the universe
has the form of a perfect fluid: $T_{\mu\nu}=(\rho
+p)U_{\mu}U_{\nu} +p g_{\mu\nu}$, where $\rho$ and $p$ are  the
energy density and pressure, respectively. We then find the
energy-supply vector
  \begin{equation}
  \Psi_a= \left (-\frac{1}{2} (\rho+p)H \tilde r, \frac{1}{2}(\rho+p)a
  \right ).
  \end{equation}
During the time internal $dt$, we obtain the amount of energy
crossing the
  apparent horizon:
  \begin{equation}
  \label{2eq12}
  -dE \equiv -A\Psi=A (\rho+p)H \tilde r_A dt,
  \end{equation}
where $A= n \Omega_n \tilde r^{n-1}_A$ is the area of the apparent
horizon. Assuming that the apparent horizon has an associated
entropy S and temperature $T$
 \begin{equation}
 \label{2eq13}
 S = \frac{A}{4G}, \ \ \  T= \frac{1}{2\pi \tilde r_A},
 \end{equation}
and then using the first law of thermodynamics, $-dE =TdS$, we
finally obtain
 \begin{equation}
 \label{2eq14}
 \dot H -\frac{k}{a^2}=-\frac{8\pi G}{n-1} (\rho +p).
 \end{equation}
Equation (\ref{2eq14}) is nothing but one of Friedmann equations
describing an $(n+1)$-dimensional FRW universe with the spatial
curvature $k$. Note that in the procedure to get Eq.
(\ref{2eq14}), we have used
  \begin{equation}
  \dot{\tilde r}_A= -H\tilde r^3_A \left (\dot H
  -\frac{k}{a^2}\right).
  \end{equation}
Once the continuity (conservation) equation of the perfect fluid
is given,
  \begin{equation}
  \label{2eq16}
  \dot \rho + n H(\rho +p) =0,
  \end{equation}
we can substitute $H(\rho +p)$ into (\ref{2eq14}),  integrate the
resulting equation, and finally obtain
  \begin{equation}
  \label{2eq17}
  H^2 +\frac{k}{a^2}= \frac{16\pi G}{n(n-1)}\rho.
  \end{equation}
This is just another Friedmann equation, the time-time component
of Einstein equations. Note that in getting the Friedmann equation
(\ref{2eq17}), an integration constant has been dropped out. In
fact, this integration constant can be regarded as a cosmological
constant, which can be incorporated into the energy density $\rho$
as a special component.  We thus have obtained the Friedmann
equations (\ref{2eq14}) and (\ref{2eq17}) for the FRW universe by
applying the first law of thermodynamics to the apparent horizon.
A passing remark is that the above procedure to obtain the
Friedmann equations from the first law of thermodynamics can still
be applied to the inflationary model with a homogenous scalar
field (inflaton), $\phi(t)$. The homogenous scalar field obeys the
field equation
\begin{equation}
\ddot{\phi} + n H \dot{\phi} + V' (\phi) = 0.
\end{equation}
Substituting the energy density and pressure
\begin{equation}
\rho = \frac{1}{2} \dot{\phi}^2 + V(\phi), \quad p = \frac{1}{2}
\dot{\phi}^2 - V(\phi),
\end{equation}
into Eq. (\ref{2eq14}) and integrating it, we can find the
Friedmann equation for the inflationary model
\begin{equation}
\label{2eq17-b} H^2 +\frac{k}{a^2}= \frac{16\pi G}{n(n-1)}
\Biggl(\frac{1}{2} \dot{\phi}^2 + V(\phi)\Biggr).
\end{equation}

Careful scrutiny of the above procedure reveals that correct
derivation of the Friedmann equations heavily depends on the
assumption given in Eq. (\ref{2eq13}): the entropy is given by a
quarter of the apparent horizon area and the temperature is
inversely proportional to the apparent horizon radius $\tilde
r_A$. The proportionality of entropy and the horizon area can be
argued by the area formula of black hole entropy and the so-called
{\it unified first law} (\ref{2eq9}). The assumption on the
temperature may be justified as follows. A direct calculation of
the surface gravity (\ref{2eq10}) at the apparent horizon gives
  \begin{eqnarray}
  \label{2eq18}
  \kappa & =& -\frac{\tilde r_A}{2}\left(\dot H +2H^2
  +\frac{k}{a^2}\right) \nonumber \\
    &=& -\frac{1}{\tilde r_A}\left(1-\frac{\dot {\tilde
    r_A}}{2H\tilde r_A}\right).
  \end{eqnarray}
For the dynamic apparent horizon, one can see that, in determining
the surface gravity, one has to know not only the apparent horizon
radius and the Hubble parameter, but also the time-dependence of
the horizon radius. Note that for a static or stationary black
hole, the surface gravity on the black hole horizon is a constant
(the zeroth law of black hole thermodynamics \cite{firstlaw}).
When the black hole mass changes by an infinitesimally small
amount, the horizon radius and thereby the Hawking temperature and
entropy will accordingly have a small change. However, the
differential form, $dM =TdS$, of the first law of black hole
thermodynamics tells us that one needs not to consider the
corresponding change of temperature in this procedure of applying
the first law of thermodynamics. Therefore, when one applies the
first law to the apparent horizon to calculate the surface gravity
and thereby the temperature and considers an infinitesimal amount
of energy crossing the apparent horizon, the apparent horizon
radius $\tilde r_A$ should be regarded to have a fixed value. In
this sense, we have $\kappa \approx -1/\tilde r_A$, and thus
recover the relation $T \equiv |\kappa|/(2\pi)=1/(2\pi \tilde
r_A)$ between the temperature and surface gravity at the apparent
horizon. In this way we justify the assumption of the temperature
in Eq. (\ref{2eq13}). In other words, the first law of
thermodynamics may hold only approximately for the dynamical
apparent horizon. This point is worth further studying.

In conclusion, employing the assumption (\ref{2eq13}) and the
first law (\ref{1eq2}) of thermodynamics to the dynamic apparent
horizon, we are able to obtain the Friedmann equations for an
$(n+1)$-dimensional FRW universe with any spatial curvature.

\sect{Friedmann equations in Gauss-Bonnet gravity}

In the previous section we have assumed that the apparent horizon
has an entropy proportional to its horizon area. This assumption
originated from the black hole thermodynamics: the entropy of
black hole horizon obeys the so-called area formula \cite{Wald}.
It is well known that the area formula of black hole entropy no
longer holds in higher derivative gravity theories. So it would be
interesting to see whether, once given a relation between the
entropy and horizon area, one can obtain or not the correct
Friedamnn equations for a gravity theory by the approach developed
in the previous section. In this section, we shall consider a
special higher derivative gravity theory$-$Gauss-Bonnet gravity.

The action of the Gauss-Bonnet gravity can be written down as
\begin{equation}
\label{3eq1} S = \frac{1}{16\pi G} \int d^{n+1}x\sqrt{-g} (R
+\alpha R_{GB})+S_m,
\end{equation}
where $\alpha$ is a constant with the dimension $[length]^{2}$,
$R_{GB}=R^2-4 R_{\mu\nu}R^{\mu\nu}+R_{\mu\nu\gamma\delta}
R^{\mu\nu\gamma\delta}$ is called the Gauss-Bonnet term and $S_m$
denotes the action of matter. The Gauss-Bonnet term naturally
appears in the low energy effective action of heterotic string
theory. The Gauss-Bonnet term is a topological term in four
dimensions, and thus does not have any dynamic effect in those
dimensions. The expansion of Gauss-Bonnet gravity around a flat
spacetime is ghost free. The Gauss-Bonnet gravity (\ref{3eq1}) is
special in the sense that although the action includes higher
derivative curvature terms, there are no more than second-order
derivative terms of metrics in equations of motion. Varying the
action, one has the equations of motion
\begin{eqnarray}
\label{3eq2} 8\pi G T_{\mu\nu} &=& R_{\mu\nu} -
\frac{1}{2}g_{\mu\nu}R
  - \alpha \Biggl( \frac{1}{2}g_{\mu\nu}R_{GB}
\nonumber\\ && -  2 RR_{\mu\nu}+4 R_{\mu\gamma}R^{\gamma}_{\ \nu}
  +4 R_{\gamma\delta}R^{\gamma\  \delta}_{\ \mu\ \ \nu}
   -2R_{\mu\gamma\delta\lambda}R_{\nu}^{\ \gamma\delta\lambda} \Biggr).
\end{eqnarray}
In the vacuum Gauss-Bonnet gravity with/without a cosmological
constant, static black hole solutions have been found and the
associated thermodynamics has been discussed (for example, see
\cite{GB}). A static, spherically symmetric black hole solution
has the metric
$$ ds^2 =-e^{\lambda(r)}dt^2 +e^{\nu(r)}dr^2 +r^2
d\Omega_{n-1}^2,$$ with
$$ e^{\lambda(r)}= e^{-\nu(r)}=1+\frac{r^2}{2\tilde\alpha}\left(1-
\sqrt{1+\frac{64\pi G\tilde \alpha
M}{n(n-1)\Omega_nr^n}}\right),$$ where $\tilde \alpha =
(n-2)(n-3)\alpha$ and $M$ is the mass of black hole. The entropy
of the black hole has the following form \cite{GB}
\begin{equation}
\label{3eq3}
 S
 =\frac{A}{4G}\left(1+\frac{n-1}{n-3}\frac{2\tilde
 \alpha}{r_+^2}\right),
 \end{equation}
where  $A= n\Omega_{n} r_+^{n-1}$ is the horizon area and $r_+$
represents the horizon radius.

Now we apply the entropy formula (\ref{3eq3}) to the apparent
horizon, assuming that the apparent horizon has the same
expression for the entropy as Eq. (\ref{3eq3}) but replacing the
black hole horizon radius $r_+$ by the apparent horizon radius
$\tilde r_A$. That is, the apparent horizon is supposed to have
the entropy
 \begin{equation}
 \label{3eq4}
 S=\frac{A}{4G}\left(1+\frac{n-1}{n-3}\frac{2\tilde
 \alpha}{\tilde r_A^2}\right),
 \end{equation}
with $A= n\Omega_{n} \tilde r_A^{n-1}$ being the apparent horizon
area. We further assume that the apparent horizon still has the
temperature $T=1/(2\pi \tilde r_A)$. This is true since the
Hawking temperature is determined by geometry itself and has
nothing to do with gravity theory explicitly. It means that once
the geometry is given, the surface gravity and then the Hawking
temperature can be determined immediately. Calculating the amount
of energy crossing the apparent horizon, which is still given by
Eq. (\ref{2eq12}), and applying the first law, $-dE=TdS$, we are
led to
 \begin{equation}
 \label{3eq5}
 \left ( 1+2 \tilde \alpha (H^2 +\frac{k}{a^2})\right)\left(\dot H
 -\frac{k}{a^2}\right)=-\frac{8\pi G}{n-1}(\rho +p).
 \end{equation}
Furthermore, substituting the continuity equation (\ref{2eq16})
into Eq. (\ref{3eq5}) and integrating the equation, we finally
obtain
 \begin{equation}
 \label{3eq6}
 H^2 +\frac{k}{a^2} +\tilde {\alpha}\left(
 H^2+\frac{k}{a^2}\right)^2 = \frac{16\pi G}{n(n-1)}\rho.
 \end{equation}
These equations (\ref{3eq5}) and (\ref{3eq6}) are nothing but the
Friedmann equations for a FRW universe in the Gauss-Bonnet gravity
given in Ref. \cite{Cai2}, where holographic entropy bounds have
been studied.

Thus, given the relation (\ref{3eq4}) between the entropy and the
horizon area in the Gauss-Bonnet gravity theory, we have indeed
been able to derive the Friedmann equations (\ref{3eq5}) and
(\ref{3eq6}) for the Gauss-Bonnet gravity by applying the first
law to the apparent horizon.

\sect{Friedmann equations in Lovelock gravity}

In this section we extend the above discussion to a more general
case, the so-called Lovelock gravity \cite{Lovelock}, which is a
generalization of the Gauss-Bonnet gravity. The Lagrangian of the
Lovelock gravity consists of the dimensionally extended Euler
densities
\begin{equation}
\label{4eq1}
 {\cal L} = \sum^m_{i=0}c_i {\cal L}_i,
 \end{equation}
 where $c_i$ is an arbitrary constant and ${\cal L}_i$ is the
 Euler density of a $(2i)$-dimensional manifold
 \begin{equation}
 \label{4eq2}
 {\cal L}_i= 2^{-i}\delta ^{a_1b_1\cdots a_ib_i}_{c_1d_1\cdots
 c_id_i} R^{c_1d_1}_{\ a_1b_1}\cdots R^{c_id_i}_{\ a_ib_i}.
 \end{equation}
Here, the generalized delta function is totally antisymmetric in
both sets of indices. ${\cal L}_0$ is set to one, therefore the
constant $c_0$ is just the cosmological constant. ${\cal L}_1$
gives us the usual curvature scalar term. In order for the general
relativity to be recovered in the low energy limit, the constant
$c_1$ has to be positive. For simplicity, we can set $c_1=1$.
${\cal L}_2$ is just the Gauss-Bonnet term discussed in the
previous section. Although the Lagrangian of the Lovelock gravity
contains higher order derivative curvature terms, there are no
terms with more than second order derivatives of metric in
equations of motion just as in the Gauss-Bonnet gravity.
Therefore, in this sense, the Lovelock theory is not a higher
derivative gravity theory.

The Lagrangian (\ref{4eq1}) looks complicated. However, static
spherically symmetric black hole solutions can be found in this
theory in the sense that a metric function is determined by
solving for a real root of a polynomial equation \cite{Whee}. More
recently, static, non-spherically symmetric black hole solutions
have been also found in the Lovelock gravity \cite{Cai3}. The
horizons of these black holes can be hypersurfaces with a
positive, zero or negative constant scalar curvature. In
particular, it has been shown that the entropy of black hole
horizon has a simple expression in terms of the horizon radius
\cite{Cai3}, while the expression for the metric function and
causal structure of these  black holes could be quite involved.
For an $(n+1)$-dimensional static,  spherically symmetric black
hole with metric~\cite{Whee,Cai3}
  \begin{equation}
  \label{4eq3}
  ds^2 =- f(r) dt^2 + f^{-1}(r)dr^2 + r^2 d\Omega_{n-1}^2,
  \end{equation}
the metric function is given by $f(r)=1-r^2 F(r)$, where $F(r)$ is
determined by solving for real roots of the following $m$th-order
polynomial equation
  \begin{equation}
  \sum^m_{i=0}\hat c_i F^i(r) = \frac{16\pi
  GM}{n(n-1)\Omega_{n}r^n}.
  \end{equation}
Here, $M$ is an integration constant, which is just the black hole
mass, and the coefficients $\hat c_i$ are given by
  \begin{eqnarray}
  \hat c_0 &=& \frac{c_0}{n(n-1)}, \ \ \hat c_1=1, \nonumber\\
  \hat c_i &=& c_i \Pi^{2m}_{j=3}(n+1-j) \ \ \ {\rm for}\ \
    i >1.
 \end{eqnarray}
In terms of the horizon radius $r_+$, the black hole entropy has
the expression \cite{Cai3}
  \begin{equation}
  \label{4eq6}
  S =\frac{A}{4G}\sum^{m}_{i=1}\frac{i(n-1)}{(n-2i+1)}\hat
  c_ir_+^{2-2i},
\end{equation}
where $A = n\Omega_{n} r_+^{n-1}$ is the horizon area of the black
hole.  Note that the cosmological constant term $c_0$ does not
enter the expression for the black hole entropy. This is
reasonable since the black hole entropy is determined by the
horizon geometry only \cite{Wald}.

Once again, we assume that the apparent horizon of the FRW
universe has an entropy of the form (\ref{4eq6}) with the black
hole horizon radius $r_+$ replaced by the apparent horizon radius
$\tilde r_A$. The temperature of the apparent horizon is still
given by $T=1/(2\pi \tilde r_A)$ and the energy crossing the
apparent horizon during the time interval $dt$ is given by Eq.
(\ref{2eq12}). Thus, the first law, $-dE=TdS$, leads us to the
equation
\begin{equation}
\label{4eq7} \sum^m_{i=1}i\hat c_i \left
(H^2+\frac{k}{a^2}\right)^{i-1}
  \left(\dot H-\frac{k}{a^2}\right) = -\frac{8\pi G}{n-1}(\rho+p).
\end{equation}
Using the continuity equation (\ref{2eq16}) and integrating Eq.
(\ref{4eq7}), we finally obtain
\begin{equation}
\label{4eq8}
 \sum^m_{i=1} \hat c_i \left(H^2 +\frac{k}{a^2}\right)^i
 =\frac{16\pi G}{n(n-1)}\rho.
 \end{equation}
Note that here the integration constant (cosmological constant)
has been included in the energy density $\rho$. These two
equations (\ref{4eq7}) and (\ref{4eq8}) are just the Friedmann
equations for a FRW universe in the Lovelock gravity \cite{DL}.
When $m=2$, equations (\ref{4eq7}) and (\ref{4eq8}) reduce to
those corresponding equations (\ref{3eq5}) and (\ref{3eq6})] of
the Gauss-Bonnet gravity.

In other words, employing the entropy form (\ref{4eq6}) for a
static, spherically symmetric black hole in the Lovelock gravity,
we have obtained correctly the Friedmann equations for a FRW
universe in the Lovelock gravity through the first law of
thermodynamics at the apparent horizon.

\sect{Conclusion and discussion}

Employing the first law of thermodynamics, $-dE=TdS$, to the
apparent horizon of a FRW universe with any spatial curvature in
arbitrary dimensions, we have derived the Friedmann equations for
the universe. Here $-dE$ denotes the amount of energy crossing the
apparent horizon during an infinitesimal time interval and
$T=1/(2\pi \tilde r_A)$ is the temperature of the apparent horizon
$\tilde r_A$. In this procedure to obtain the Friedmann equations,
the entropy $S$ is assumed to be a quarter of the apparent horizon
area. Using the same form for the relation between entropy and
horizon area as that of a static, spherical symmetric black hole
in Gauss-Bonnet gravity and its generalization, Lovelock gravity,
we have also successfully obtained the Friedmann equations in
those gravity theories. These results are certainly related to the
holographic properties of gravity. It would be of great interest
to study further the implications of these observations to the
holographic principle.

The first law of thermodynamics is certainly valid for the
cosmological event horizon \cite{GH,Cai1}. It is not a priori
clear whether it still holds for a dynamic apparent horizon. Our
results in turn also indicate that the first law indeed holds at
the apparent horizon. Although the cosmological event horizon does
not always exist, an interesting question may arise: whether can
we still derive the Friedmann equations by applying the first law
of thermodynamics to the cosmological event horizon?  Assuming
that the event horizon has the temperature and entropy given by
Eq. (\ref{1eq3}), which are exactly correct for a de Sitter
universe, and after a naive calculation, we have found that we can
derive the Friedmann equations only for a FRW universe with $k=0$,
that is, for a flat universe. This can be understood since $\tilde
r_A = 1/H$ for $k = 0$ and the assumptions (\ref{1eq3}) and
(\ref{1eq4}) thus become equivalent to each other when $k=0$. The
failure to obtain the Friedmann equations may due to the fact that
the assumption (\ref{1eq3}) is not suitable for the cosmological
event horizon. In other words, the thermodynamic relations in Eq.
(\ref{1eq3}) may be not correct for the cosmological event
horizon. The failure may also be due to another fact that the
Friedmann equations describe local properties of spacetimes, while
the cosmological event horizon is determined by global properties
of spacetimes [see (\ref{2eq4})]. That is, one cannot determine
local properties at each point in a spacetime using global
properties of the spacetime, of course, although the inverse is
certainly true. We should point out that here the apparent horizon
is determined locally.  This may be just the reason behind the
difference between the cosmological event horizon and the apparent
horizon. This is a quite interesting issue worthy of a further
study.

Given an entropy relation to the apparent horizon area, we have
obtained the correct Friedmann equations for a FRW universe in the
Gauss-Bonnet gravity and Lovelock gravity. However, we are not
sure whether it is always true for any other gravity theory that,
assuming a relation between entropy and its horizon area, the
Friedmann equations can be obtained in that gravity theory, where
the relation of entropy and area arises. For example, for the
gravity theory of the form ${\cal L} = f(R)$, where $f$ is a
function of scalar curvature $R$, we know that the entropy of a
static black hole is $S=Af'(R)|_{r=r_+}/4G$. Can one derive the
Friedmann equations for the gravity theory? These issues are
currently under investigation.

\section*{Acknowledgments}
We thank  Seoktae Koh for useful discussions.  RGC would like to
appreciate the warm hospitality of Kunsan National University. The
work of RGC was supported in part by a grant from Chinese Academy
of Sciences, by NSFC under grants No. 10325525 and No. 90403029,
and by the Ministry of Science and Technology of China under grant
No. TG1999075401. The work of SPK was support by Korea Astronomy
Observatory.


\end{document}